# An approach based on distributed dislocations and disclinations for crack problems in couple-stress elasticity

P.A. Gourgiotis , H.G. Georgiadis *

*Mechanics Division, National Technical University of Athens,*
*Zographou Campus, Zographou, GR-15773, Greece*

**Abstract**

The technique of distributed dislocations proved to be in the past an effective approach in studying crack problems within classical elasticity. The present work is intended to extend this technique in studying crack problems within couple-stress elasticity, i.e. within a theory accounting for effects of microstructure. This extension is not an obvious one since rotations and couple-stresses are involved in the theory employed to analyze the crack problems. Here, the technique is introduced to study the case of a mode I crack. Due to the nature of the boundary conditions that arise in couple-stress elasticity, the crack is modeled by a continuous distribution of climb dislocations and constrained wedge disclinations (the concept of 'constrained wedge disclination' is first introduced in the present work). These distributions create both standard stresses and couple stresses in the body. In particular, it is shown that the mode-I case is governed by a system of coupled singular integral equations with both Cauchy-type and logarithmic kernels. The numerical solution of this system shows that a cracked solid governed by couple-stress elasticity behaves in a more rigid way (having increased stiffness) as compared to a solid governed by classical elasticity. Also, the stress level at the crack-tip region is appreciably higher than the one predicted by classical elasticity.

**Keywords:** Cracks; Dislocations; Disclinations; Couple-stress elasticity; Singular integral equations.

---------------------------------------------------------------------------------

\* Corresponding author. Tel.: +30 210 7721365; fax: +30 210 7721302.
  *E-mail address*: georgiad@central.ntua.gr (H.G. Georgiadis).

# 1. Introduction

The present work introduces an approach based on distributed dislocations and disclinations (and associated singular integral equations) to deal with the mode I crack problem of couple-stress elasticity. This theory assumes that, within an elastic body, the surfaces of each material element are subjected not only to normal and tangential forces but also to moments per unit area. The latter are called couple-stresses. Such an assumption is appropriate for materials with granular or crystalline structure, where the interaction between adjacent elements may introduce internal moments. In this way, characteristic material lengths appear representing microstructure. As is well-known, the fundamental concepts of the couple-stress theory were first introduced by Voigt (1887) and the Cosserat brothers (1909), but the subject was generalized and reached maturity only in the 1960s through the works of Toupin (1962), Mindlin and Tiersten (1962), and Koiter (1964).

The theory of couple-stress elasticity assumes that: (i) each material particle has three degrees of freedom, (ii) an augmented form of the Euler-Cauchy principle with a non-vanishing couple traction prevails, and (iii) the strain-energy density depends upon both strain and the gradient of rotation. The theory is different from the Cosserat (or micropolar) theory that takes material particles with six independent degrees of freedom (three displacement components and three rotation components, the latter involving rotation of a micro-medium w.r.t. its surrounding medium). Sometimes, the name 'restricted Cosserat theory' appears in the literature for the couple-stress theory.

It is noted that couple-stress elasticity had already in the 1960s some successful application on stress-concentration problems concerning holes and inclusions (see e.g. Mindlin, 1963; Weitsman, 1965; Bogy and Sternberg, 1967a, b; Hsu et al., 1972; Takeuti et al., 1973). In recent years, there is a renewed interest in couple-stress theory (and related generalized continuum theories) dealing with problems of microstructured materials. For instance, problems of dislocations, plasticity, fracture and wave propagation have been analyzed within the framework of couple-stress theory. This is due to the inability of the classical theory to predict the experimentally observed size effect and also due to the increasing demands for manufacturing devices at very small scales. Recent applications include work by, among others, Fleck et al. (1994), Vardoulakis and Sulem (1995), Huang et al. (1997; 1999), Fleck and Hutchinson (1998), Zhang et al. (1998), Anthoine (2000), Lubarda and Markenscoff (2000), Bardet and Vardoulakis (2001), Georgiadis and Velgaki (2003), Lubarda (2003), Ravi Shankar et al. (2004), Grentzelou and Georgiadis (2005), and Radi (2007).

Generally, the couple-stress theory is intended to model situations where the material is deformed in *very small* volumes, such as in the immediate vicinity of crack tips, notches, small holes and inclusions, and micrometer indentations. Examples of successful modelling of microstructure and size effects by this theory are provided by Kakunai et al. (1985) and Lakes (1995), among others. Also, a recent work by Bigoni and Drugan (2007) provides additional references and an interesting account of the determination of moduli via homogenization of heterogeneous materials.

Regarding now crack problems, there is a limited number of studies concerning such problems in couple-stress theory. Sternberg and Muki (1967) were the first to study the mode I finite-length crack elasticity problem by employing the method of dual integral equations. In their work, only asymptotic results were obtained showing that both the stress and couple-stress fields exhibit a square-root singularity, while the rotation field is bounded at the crack-tip. Adopting the same method, Ejike (1969) studied the problem of a circular (penny-shaped) crack in couple-stress elasticity. Later, Atkinson and Leppington (1977) studied the problem of a semi-infinite crack by using the Wiener-Hopf technique. More recently, Huang et al. (1997) using the method of eigenfunction expansions, provided near-tip asymptotic fields for mode I and mode II crack problems in couple-stress elasticity. Also, Huang et al. (1999) using the Wiener-Hopf technique obtained full-field solutions for semi-infinite cracks under in-plane loading in elastic-plastic materials with strain-gradient effects of the couple-stress type.

The aim of the present investigation is to extend the distributed dislocation technique (and the related singular integral equation technique) in dealing with crack problems of couple-stress elasticity and to obtain, for the first time, a *full-field* solution to the mode I problem of a *finite-length* crack. The couple-stress case is our first attempt to introduce singular integral equations in crack problems of generalized continua. Efforts dealing with gradient elasticity are also under way. Here, we introduce an approach based on distributed dislocations and disclinations. In particular, the concept of a special type of disclination (we call it 'constrained wedge disclination') is employed in order to deal with the features of the couple-stress theory. No such concept was needed in dealing with crack problems within the classical elasticity theory. For the latter problems, the standard distributed dislocation technique (DDT) was introduced by Bilby et al. (1963, 1968). This is an analytical/numerical technique. The strength of the DDT lies in the fact that it gives detailed full-field solutions for crack problems at the expense of relatively little analytical and computational demands as compared to the elaborate analytical method of dual integral equations or the standard numerical methods of Finite and Boundary Elements. A thorough exposition of the technique and the treatment of various crack problems can be found in the treatise by Hills et al. (1996).

Despite the numerous applications of the DDT in classical elasticity, it appears that there is a limited work in solving crack problems with this technique in materials with microstructure. Recently, the present authors (Gourgiotis and Georgiadis, 2007) applied the standard DDT to solve finite-length crack problems, under mode II and mode III conditions, within the framework of couple-stress elasticity. Within this framework, and having solved now the mode I (opening mode) case, a comparison between the two plane-strain crack modes (mode I and mode II) shows that mode I is mathematically more involved than mode II. Certainly, this is in contrast with situations of classical elasticity where solving problems of mode I and mode II involves the same mathematical effort. The additional effort in dealing with the mode I case here is due to the nature of the boundary conditions that arise in couple-stress elasticity (involving rotations and couple-stresses). However, such a situation does not appear in the mode II case of couple-stress elasticity (Gourgiotis and Georgiadis, 2007).

As in analogous situations of classical elasticity, a superposition scheme will be followed. Thus, the solution to the basic problem (body with a traction-free crack under a remote constant tension) will be obtained by the superposition of the stress and couple-stress fields arising in an un-cracked body (of the same geometry) to the 'corrective' stresses and couple-stresses induced by a distribution of defects chosen so that the crack-faces become traction-free. Due to the nature of the boundary conditions, it will be shown that in order to obtain the corrective solution, we need to distribute not only climb dislocations but also constant discontinuities of the rotation along the crack faces. We name the latter discontinuities *constrained wedge disclinations*. The term 'constrained' refers to the requirement of *zero* normal displacement along the disclination plane. Notice that according to the standard notion of a wedge disclination (see e.g. Anthony, 1970; de Wit, 1973), the normal displacement is also discontinuous along the disclination plane and increases linearly with distance from the core becoming unbounded at infinity. Clearly, a standard wedge disclination would not serve our purpose here. The concept of a constrained wedge disclination is first introduced in the present work (see Sections 4 and 5 below for the details).

The Green's functions of our problem (i.e. the stress fields due to a discrete climb dislocation and a discrete constrained wedge disclination) are obtained by the use of Fourier transforms. Finally, it is shown that the continuous distribution of the discontinuities along the crack faces results in a system of coupled singular integral equations with both Cauchy-type and logarithmic kernels. The numerical solution of this system shows that a cracked solid governed by couple-stress elasticity behaves in a more rigid way (having increased stiffness) as compared to a solid governed by classical elasticity. Also, the stress level at the crack-tip region is appreciably higher than the one predicted by classical elasticity.

## 2. Fundamentals of couple-stress elasticity

In this Section, the basic equations of couple-stress elasticity are briefly presented. As mentioned before, couple-stress elasticity assumes that: (i) each material particle has three degrees of freedom, (ii) an augmented form of the Euler-Cauchy principle with a non-vanishing couple traction prevails, and (iii) the strain-energy density depends upon both strain and the gradient of rotation.

In addition to the fundamental papers by Mindlin and Tiersten (1962) and Koiter (1964), interesting presentations of the theory can be found in the works by Aero and Kuvshinskii (1960), Palmov (1964), and Muki and Sternberg (1965). The basic equations of dynamical couple-stress theory (including the effects of micro-inertia) were given by Georgiadis and Velgaki (2003).

In the absence of inertia effects, for a control volume $CV$ with bounding surface $S$, the balance laws for the linear and angular momentum read

$$\int_S T_i^{(n)} dS + \int_{CV} F_i \, d(CV) = 0 \quad , \tag{1}$$

$$\int_S \left( x_j T_k^{(n)} e_{ijk} + M_i^{(n)} \right) dS + \int_{CV} \left( x_j F_k e_{ijk} + C_i \right) d(CV) = 0 \quad , \tag{2}$$

where a Cartesian rectangular coordinate system $Ox_1 x_2 x_3$ is used along with indicial notation and summation convention, $e_{ijk}$ is the Levi-Civita alternating symbol, **n** is the outward unit vector normal to the surface with direction cosines $n_j$, $T_i^{(n)}$ is the surface force per unit area (force traction), $F_i$ is the body force per unit volume, $M_i^{(n)}$ is the surface moment per unit area (couple traction), and $C_i$ is the body moment per unit volume.

Next, pertinent *force-stress* and *couple-stress* tensors are introduced by considering the equilibrium of the elementary material tetrahedron and enforcing (1) and (2), respectively. The force stress or total stress tensor $\sigma_{ij}$ (which is asymmetric) is defined by

$$T_i^{(n)} = \sigma_{ji} n_j \quad , \tag{3}$$

and the couple-stress tensor $\mu_{ij}$ (which is also asymmetric) by

$$M_i^{(n)} = \mu_{ji} n_j \ . \tag{4}$$

In addition, just like the third Newton's law $\mathbf{T}^{(\mathbf{n})} = -\mathbf{T}^{(-\mathbf{n})}$ is proved to hold by considering the equilibrium of a material 'slice', it can also be proved that $\mathbf{M}^{(\mathbf{n})} = -\mathbf{M}^{(-\mathbf{n})}$ (see e.g. Jaunzemis, 1967). The couple-stresses $\mu_{ij}$ are expressed in dimensions of [force][length]$^{-1}$. Further, $\sigma_{ij}$ can be decomposed into a symmetric and anti-symmetric part

$$\sigma_{ij} = \tau_{ij} + \alpha_{ij} \ , \tag{5}$$

with $\tau_{ij} = \tau_{ji}$ and $\alpha_{ij} = -\alpha_{ji}$, whereas it is advantageous to decompose $\mu_{ij}$ into its deviatoric $\mu_{ij}^{(D)}$ and spherical $\mu_{ij}^{(S)}$ part in the following manner

$$\mu_{ij} = m_{ij} + \frac{1}{3}\delta_{ij}\mu_{kk} \ , \tag{6}$$

where $m_{ij} = \mu_{ij}^{(D)}$, $\mu_{ij}^{(S)} = (1/3)\delta_{ij}\mu_{kk}$, and $\delta_{ij}$ is the Kronecker delta. Now, with the above definitions and the help of the Green-Gauss theorem, one may obtain the stress equations of motion. Equation (2) leads to the following moment equation

$$\partial_i \mu_{ij} + \sigma_{ki} e_{ijk} + C_j = 0 \ , \tag{7}$$

which can also be written as

$$\frac{1}{2}\partial_i \mu_{il} e_{jkl} + \alpha_{jk} + \frac{1}{2}C_l e_{jkl} = 0 \ , \tag{8}$$

since by its definition the anti-symmetric part of stress is written as $\boldsymbol{\alpha} \equiv -(1/2)\mathbf{I} \times (\boldsymbol{\sigma} \times \mathbf{I})$, where $\mathbf{I}$ is the idemfactor. Also, Eq. (1) leads to the following force equation

$$\partial_j \sigma_{jk} + F_k = 0 \ , \tag{9}$$

or, by virtue of (5), to the equation

$$\partial_j \tau_{jk} + \partial_j \alpha_{jk} + F_k = 0 \ . \tag{10}$$

Further, combining (8) and (10) yields the single equation

$$\partial_j \tau_{jk} - \frac{1}{2}\partial_j \partial_i \mu_{il} e_{jkl} + F_k - \frac{1}{2}\partial_j C_l e_{jkl} = 0 \ . \tag{11}$$

Finally, in view of Eq.(6) and by taking into account that $\mathrm{curl}\big(\mathrm{div}\big((1/3)\delta_{ij}\mu_{kk}\big)\big) = 0$, we write (11) as

$$\partial_j \tau_{jk} - \frac{1}{2}\partial_j \partial_i m_{il} e_{jkl} + F_k - \frac{1}{2}\partial_j C_l e_{jkl} = 0 \ , \tag{12}$$

which is the final *equation of equilibrium*.

Now, as for the kinematical description of the continuum, the following quantities are defined in the framework of the geometrically linear theory

$$\varepsilon_{ij} = \frac{1}{2}\big(\partial_j u_i + \partial_i u_j\big) \ , \tag{13}$$

$$\omega_{ij} = \frac{1}{2}\big(\partial_j u_i - \partial_i u_j\big) \ , \tag{14}$$

$$\omega_i = \frac{1}{2}e_{ijk}\partial_j u_k \ , \tag{15}$$

$$\kappa_{ij} = \partial_i \omega_j \ , \tag{16}$$

where $\varepsilon_{ij}$ is the strain tensor, $\omega_{ij}$ is the rotation tensor, $\omega_i$ is the rotation vector, and $\kappa_{ij}$ is the curvature tensor (i.e. the gradient of rotation or the curl of the strain) expressed in dimensions of [length]$^{-1}$. Notice also that Eq. (16) can alternatively be written as

$$\kappa_{ij} = \frac{1}{2} e_{jkl} \partial_i \partial_k u_l = e_{jkl} \partial_k \varepsilon_{il} \quad . \tag{17}$$

Equation (17) expresses compatibility for curvature and strain fields. In addition, there is an identity $\partial_k \kappa_{ij} = \partial_i \partial_k \omega_j = \partial_i \kappa_{kj}$, which defines the compatibility equations for the curvature components. The compatibility equations for the strain components are the usual Saint Venant's compatibility equations (see e.g. Jaunzemis, 1967). We notice also that $\kappa_{ii} = 0$ because $\kappa_{ii} = \partial_i \omega_i = (1/2) e_{ijk} u_{k,ji} = 0$ and, therefore, $\kappa_{ij}$ has only eight independent components. The tensor $\kappa_{ij}$ is obviously an *asymmetric* tensor.

Regarding traction boundary conditions, at any point on a smooth boundary or section, the following three *reduced* force-tractions and two *tangential* couple-tractions should be specified (Mindlin and Tiersten, 1962; Koiter, 1964)

$$P_i^{(n)} = \sigma_{ji} n_j - \frac{1}{2} e_{ijk} n_j \partial_k m_{(nn)} , \tag{18}$$

$$R_i^{(n)} = m_{ji} n_j - m_{(nn)} n_i , \tag{19}$$

where $m_{(nn)} = n_i n_j m_{ij}$ is the normal component of the deviatoric couple-stress tensor $m_{ij}$. The modifications for the case in which corners appear along the boundary can be found in the article by Koiter (1964).

It is worth noticing that at first sight, it might seem plausible that the surface tractions (i.e. the force-traction and the couple-traction) can be prescribed arbitrarily on the external surface of the body through relations (3) and (4), which stem from the equilibrium of the material tetrahedron. However, as Koiter (1964) pointed out, the resulting number of six traction boundary conditions (three force-tractions and three couple-tractions) would be in contrast with the *five* geometric boundary conditions that can be imposed. Indeed, since the rotation vector $\omega_i$ in couple-stress elasticity is not independent of the displacement vector $u_i$ (as (15) suggests), the normal component of the rotation is fully specified by the distribution of tangential displacements over the boundary. Therefore, only the three displacement and the two tangential rotation components can be prescribed independently. As a consequence, only *five* surface tractions (i.e. the work conjugates of the above five independent kinematical quantities) can be specified at a point of the bounding surface of the body, i.e. Eqs. (18) and (19). On the contrary, in the Cosserat (micropolar) theory, the traction boundary conditions are six since the rotation is fully independent of the displacement vector (see e.g. Nowacki, 1972). In the latter case, the tractions can directly be derived from the equilibrium of the material tetrahedron, so (3) and (4) are the pertinent traction boundary conditions.

Introducing the constitutive equations of the theory is now in order. We assume a linear and isotropic material response, in which case the strain-energy density takes the form

$$W \equiv W(\varepsilon_{ij}, \kappa_{ij}) = \frac{1}{2}\lambda \varepsilon_{ii}\varepsilon_{jj} + \mu \varepsilon_{ij}\varepsilon_{ij} + 2\eta \kappa_{ij}\kappa_{ij} + 2\eta' \kappa_{ij}\kappa_{ji} \ , \tag{20}$$

where $(\lambda, \mu, \eta, \eta')$ are material constants. Then, Eq. (20) leads, through the standard variational manner, to the following constitutive equations

$$\tau_{ij} \equiv \sigma_{(ij)} = \frac{\partial W}{\partial \varepsilon_{ij}} = \lambda \delta_{ij}\varepsilon_{kk} + 2\mu \varepsilon_{ij} \ , \tag{21}$$

$$m_{ij} = \frac{\partial W}{\partial \kappa_{ij}} = 4\eta \kappa_{ij} + 4\eta' \kappa_{ji} \ . \tag{22}$$

In view of (21) and (22), the moduli $(\lambda, \mu)$ have the same meaning as the Lamé constants of classical elasticity theory and are expressed in dimensions of [force][length]$^{-2}$, whereas the moduli $(\eta, \eta')$ account for couple-stress effects and are expressed in dimensions of [force].

Next, incorporating the constitutive relations (21) and (22) into the equation of equilibrium (12) and using the geometric relations (13)-(16), one may obtain the equations of equilibrium in terms of displacement components (Muki and Sternberg, 1965), i.e.

$$\nabla^2 u_i - \ell^2 \nabla^4 u_i + \partial_i \left( \frac{1}{1-2\nu}(\nabla \cdot \mathbf{u}) + \ell^2 \nabla^2 (\nabla \cdot \mathbf{u}) \right) = 0 \ , \tag{23}$$

where $\nu$ is Poisson's ratio, $\ell \equiv (\eta/\mu)^{1/2}$ is a characteristic material length, and the absence of body forces and couples is assumed. In the limit $\ell \to 0$, the Navier-Cauchy equations of classical linear isotropic elasticity are recovered from (23). Indeed, the fact that Eqs. (23) have an increased order w.r.t. their limit case (recall that the Navier-Cauchy equations are PDEs of the second order) and the coefficient $\ell$ multiplies the higher-order term reveals the *singular-perturbation* character of the couple-stress theory and the emergence of associated *boundary-layer* effects.

Finally, the following points are of notice: (i) Since $\kappa_{ii} = 0$, $m_{ii} = 0$ is also valid and therefore the tensor $m_{ij}$ has only eight independent components. (ii) The scalar $(1/3)\mu_{kk}$ of the couple-stress tensor does not appear in the final equation of equilibrium, nor in the reduced boundary conditions and the constitutive equations. Consequently, $(1/3)\mu_{kk}$ is left indeterminate within the couple-stress theory. (iii) The following restrictions for the material constants should prevail on the basis of a positive definite strain-energy density (Mindlin and Tiersten, 1962)

$$3\lambda + 2\mu > 0 , \quad \mu > 0 , \quad \eta > 0 , \quad -1 < \frac{\eta'}{\eta} < 1 . \tag{24a-d}$$

## 3. Basic equations in plane-strain

For a body that occupies a domain in the $(x, y)$-plane under conditions of plane strain, the displacement field takes the general form

$$u_x \equiv u_x(x, y) \neq 0 , \quad u_y \equiv u_y(x, y) \neq 0 , \quad u_z \equiv 0 . \tag{25a-c}$$

First, the components of the force-stress and couple-stress tensors will be obtained. The independence upon the coordinate $z$ of *all* components of the force-stress and couple-stress tensors, under the assumption (25c), was proved by Muki and Sternberg (1965). Indeed, it is noteworthy that, contrary to the respective plane-strain case in the conventional theory, this independence is not obvious within the couple-stress theory. Notice further that except for $\omega_z \equiv \omega$ and $(\kappa_{xz}, \kappa_{yz})$ all others components of the rotation vector and the curvature tensor vanish identically in the particular case of plane-strain considered here. The non-vanishing components $(\tau_{xx}, \tau_{xy}, \tau_{yy})$ and $(m_{xz}, m_{yz})$ follow from (21) and (22), respectively. Then, $(\alpha_{xx}, \alpha_{xy}, \alpha_{yx}, \alpha_{yy})$ are found from (8) and, finally, $(\sigma_{xx}, \sigma_{xy}, \sigma_{yx}, \sigma_{yy})$ are provided by (5). Vanishing body forces and body couples are assumed in what follows. In view of the above, the following expressions are written

$$m_{xz} = 2\mu\ell^2 \left( \frac{\partial^2 u_y}{\partial x^2} - \frac{\partial^2 u_x}{\partial x \partial y} \right) , \tag{26}$$

$$m_{yz} = 2\mu\ell^2\left(\frac{\partial^2 u_y}{\partial x \partial y} - \frac{\partial^2 u_x}{\partial y^2}\right), \tag{27}$$

$$\alpha_{xx} = \alpha_{yy} = 0, \tag{28}$$

$$\alpha_{yx} = \frac{1}{2}\left(\frac{\partial m_{xz}}{\partial x} + \frac{\partial m_{yz}}{\partial y}\right), \tag{29}$$

$$\alpha_{xy} = -\alpha_{yx}, \tag{30}$$

$$\sigma_{xx} = (\lambda + 2\mu)\frac{\partial u_x}{\partial x} + \lambda\frac{\partial u_y}{\partial y}, \tag{31}$$

$$\sigma_{yy} = (\lambda + 2\mu)\frac{\partial u_y}{\partial y} + \lambda\frac{\partial u_x}{\partial x}, \tag{32}$$

$$\sigma_{yx} = \mu\left(\frac{\partial u_x}{\partial y} + \frac{\partial u_y}{\partial x}\right) + \mu\ell^2\left(\frac{\partial^3 u_y}{\partial x^3} - \frac{\partial^3 u_x}{\partial x^2 \partial y} + \frac{\partial^3 u_y}{\partial x \partial y^2} - \frac{\partial^3 u_x}{\partial y^3}\right), \tag{33}$$

$$\sigma_{xy} = \mu\left(\frac{\partial u_x}{\partial y} + \frac{\partial u_y}{\partial x}\right) - \mu\ell^2\left(\frac{\partial^3 u_y}{\partial x^3} - \frac{\partial^3 u_x}{\partial x^2 \partial y} + \frac{\partial^3 u_y}{\partial x \partial y^2} - \frac{\partial^3 u_x}{\partial y^3}\right). \tag{34}$$

Incorporating (25a-c) into the equations of equilibrium in (23), we obtain the following system of coupled PDEs of the fourth order for the displacement components $(u_x, u_y)$

$$\frac{1}{1-2\nu}\frac{\partial}{\partial x}\left(2(1-\nu)\frac{\partial u_x}{\partial x} + \frac{\partial u_y}{\partial y}\right) + \frac{\partial^2 u_x}{\partial y^2} + \ell^2\left(\frac{\partial^4 u_y}{\partial x^3 \partial y} - \frac{\partial^4 u_x}{\partial x^2 \partial y^2} + \frac{\partial^4 u_y}{\partial x \partial y^3} - \frac{\partial^4 u_x}{\partial y^4}\right) = 0, \tag{35}$$

$$\frac{1}{1-2\nu}\frac{\partial}{\partial y}\left(2(1-\nu)\frac{\partial u_y}{\partial y} + \frac{\partial u_x}{\partial x}\right) + \frac{\partial^2 u_y}{\partial x^2} + \ell^2\left(\frac{\partial^4 u_x}{\partial x^3 \partial y} - \frac{\partial^4 u_y}{\partial x^2 \partial y^2} + \frac{\partial^4 u_x}{\partial x \partial y^3} - \frac{\partial^4 u_y}{\partial x^4}\right) = 0. \tag{36}$$

## 4. Formulation of the crack problem

Consider a straight crack of finite length $2a$ embedded in a body of infinite extent in the $xy$-plane (Fig. 1). The body is governed by the equations of couple-stress elasticity and it is in a field of uniform uni-axial tension, under plane-strain conditions. The crack faces are traction free and are defined by $\mathbf{n} = (0, \pm 1)$. Then, according to (18) and (19), the boundary conditions along the crack faces are written as

$$\sigma_{yx}(x,0)=0\,,\quad \sigma_{yy}(x,0)=0\,,\quad m_{yz}(x,0)=0 \qquad \text{for } |x|<a\,. \tag{37a-c}$$

The regularity conditions at infinity are

$$\sigma_{yx}^{\infty},\sigma_{xy}^{\infty},\sigma_{xx}^{\infty} \to 0\,,\quad \sigma_{yy}^{\infty} \to \sigma_0\,,\quad m_{xz}^{\infty}, m_{yz}^{\infty} \to 0\,,\quad \text{as } r \to \infty\,, \tag{38a-c}$$

where $r=(x^2+y^2)^{1/2}$ is the distance from the origin, and the constant $\sigma_0$ denotes the remotely applied normal loading.

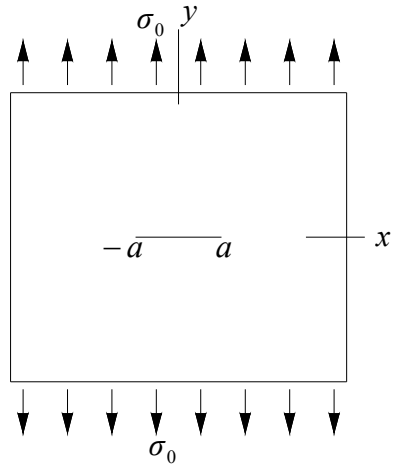

Fig. 1 Cracked body under remote tension field in plane strain.

Now, the crack problem is decomposed into the following two auxiliary problems.

*The un-cracked body*

The displacement and the rotation field for the un-cracked body problem are given as (Sternberg and Muki, 1967)

$$u_x = -\frac{v\sigma_0}{2\mu}x \ , \quad u_y = \frac{(1-v)\sigma_0}{2\mu}y \ , \quad \omega = 0 \ . \tag{39}$$

The stress field can readily be obtained from (26)-(34) as

$$\sigma_{yy}(x,y) = \sigma_0 \ , \quad \sigma_{xx} = \sigma_{yx} = \sigma_{xy} = 0 \ , \quad m_{xz} = m_{yz} = 0 \ . \tag{40a-c}$$

Notice that there are no couple-stresses induced in the un-cracked body, the body being in a state of pure tension.

*The corrective solution*

Consider next a body geometrically identical to the initial cracked body (Fig. 1) but with no remote loading now. The only loading applied is along the crack faces. This consists of equal and opposite tractions to those generated in the un-cracked body. The boundary conditions along the faces of the crack are written as

$$\sigma_{yy}(x,0) = -\sigma_0 \ , \quad m_{yz}(x,0) = 0 \ , \quad \sigma_{yx}(x,0) = 0 \quad \text{for} \quad |x| < a \ . \tag{41a-c}$$

Notice that in classical elasticity it would suffice a continuous distribution of *climb* dislocations with Burger's vector $\mathbf{b} = (0,b,0)$ to produce the desired normal stresses (41a). However, this is not the case in couple-stress elasticity because a discrete climb dislocation produces *both* normal stresses $\sigma_{yy}$ and couple-stresses $m_{yz}$ along the dislocation line $y = 0$. Therefore, it is not possible to satisfy both (41a) and (41b) only by a continuous distribution of climb dislocations. On the other hand, within the framework of couple-stress elasticity, we know that the work conjugates of the reduced force traction $P_y = \sigma_{yy}n_y$ and the tangential couple traction $R_z = m_{yz}n_y$ are the normal displacement $u_y$ and the rotation $\omega$, respectively. In light of the above, we are led to the conclusion that in order to satisfy all the boundary conditions in (41) we should distribute discontinuities of *both* displacement $u_y$ (i.e. climb dislocations) and rotation $\omega$ (the so-called constrained wedge disclinations) along the crack faces.

It is noteworthy that in the mode II crack problem of couple-stress elasticity studied by the present authors (Gourgiotis and Georgiadis, 2007), only a distribution of glide dislocations was indeed sufficient to generate the requisite shear stress $\sigma_{yx}$ along the crack-faces. This is because a discrete glide dislocation produces neither normal stresses $\sigma_{yy}$ nor couple-stresses $m_{yz}$ along the crack-line $y = 0$. In that problem, employing the standard DDT was sufficient and led to a *single* singular integral equation. On the contrary, in the present mode I crack problem, the distribution of both climb dislocations and constrained wedge disclinations leads to a *system* of coupled singular integral equations for the dislocation and the disclination densities.

Our next aim is to determine the stress and couple-stress fields induced by a discrete climb dislocation and a discrete constrained wedge disclination. Both defects are located at the origin of the $(x, y)$-plane. These stress fields will serve as the Green's functions for our crack problem.

## 5. Green's functions (climb dislocation and constrained wedge disclination)

Due to the symmetry of both problems w.r.t. the plane $y = 0$, only the upper half-plane domain ($-\infty < x < \infty$, $y \geq 0$) will be considered. In this domain, the Fourier transform is utilized to suppress the $x$-dependence in the field equations and the boundary conditions. The direct Fourier transform and its inverse are defined as follows

$$f^*(\xi, y) = \frac{1}{(2\pi)^{1/2}} \int_{-\infty}^{\infty} f(x, y) e^{ix\xi} dx \,, \tag{42a}$$

$$f(x, y) = \frac{1}{(2\pi)^{1/2}} \int_{-\infty}^{\infty} f^*(\xi, y) e^{-ix\xi} d\xi \,, \tag{42b}$$

where $i \equiv (-1)^{1/2}$. Transforming now (35) and (36) with (42a) gives a system of ordinary differential equations for $(u_x^*, u_y^*)$ written in the following compact form

$$[K] \begin{bmatrix} u_x^* \\ u_y^* \end{bmatrix} = \begin{bmatrix} 0 \\ 0 \end{bmatrix} \,, \tag{43}$$

where the differential operator $[K]$ is given as

$$[K] = \begin{bmatrix} -\ell^2 d^4 + (1+\ell^2\xi^2)d^2 - (1+k)\xi^2 & i\xi\ell^2 d(\xi^2 - d^2) - i\xi kd \\ i\xi\ell^2 d(\xi^2 - d^2) - i\xi kd & (1+\xi^2\ell^2 + k)d^2 - \xi^2(1+\xi^2\ell^2) \end{bmatrix}, \quad (44)$$

with $k = 1/(1-2\nu)$, $d(\ ) \equiv d(\ )/dy$, $d^2(\ ) \equiv d^2(\ )/dy^2$, etc.

The system of homogeneous differential equations in (43) has a solution different than the trivial one if and only if the determinant of $[K]$ is zero. Hence,

$$(d^2 - \xi^2)^2 [\ell^2(d^2 - \xi^2) - 1] = 0. \quad (45)$$

The latter equation has two double roots $d = \pm|\xi|$ and two single roots $d = \pm(1+\ell^2\xi^2)^{1/2}/\ell$. The first pair is the same as in classical elasticity, whereas the second pair reflects the presence of couple-stress effects. The general solution of (43) is obtained after some rather extensive algebra and it has the following form that is bounded as $y \to +\infty$

$$u_x^*(\xi, y) = A_1(\xi)e^{-|\xi|y} + A_2(\xi)ye^{-|\xi|y} + A_3(\xi)e^{-\frac{y\alpha}{\ell}}, \quad (46)$$

$$u_y^*(\xi, y) = -i\xi^{-1}\left[|\xi|A_1(\xi) + (3-4\nu)A_2(\xi)\right]e^{-|\xi|y} - iy\,\text{sgn}(\xi)A_2(\xi)e^{-|\xi|y}$$

$$-i\frac{\xi\ell}{\alpha}A_3(\xi)e^{-\frac{y\alpha}{\ell}}, \quad (47)$$

Where $\alpha \equiv \alpha(\xi) = (1+\ell^2\xi^2)^{1/2}$, sgn( ) is the signum function, and the functions $(A_1(\xi), A_2(\xi), A_3(\xi))$ will be determined through the enforcement of boundary conditions in each specific problem.

Having in hand the transformed general solution (46) and (47), the transformed rotation, stresses and couple-stresses may follow by the use of the following expressions

$$\omega^*(\xi, y) = -\frac{1}{2}\left(i\xi u_y^* + \frac{du_x^*}{dy}\right), \quad (48)$$

$$\sigma_{yy}^*(\xi, y) = \frac{2\mu}{1-2v}\left[(1-v)\frac{du_y^*}{dy} - iv\xi u_x^*\right], \tag{49}$$

$$\sigma_{xx}^*(\xi, y) = \frac{2\mu}{1-2v}\left[-(1-v)i\xi u_x^* + v\frac{du_y^*}{dy}\right], \tag{50}$$

$$\sigma_{yx}^*(\xi, y) = \mu\left(-i\xi u_y^* + \frac{du_x^*}{dy}\right) + \mu\ell^2\left(i\xi^3 u_y^* + \xi^2\frac{du_x^*}{dy} - i\xi\frac{d^2 u_y^*}{dy^2} - \frac{d^3 u_x^*}{dy^3}\right), \tag{51}$$

$$\sigma_{xy}^*(\xi, y) = \mu\left(-i\xi u_y^* + \frac{du_x^*}{dy}\right) - \mu\ell^2\left(i\xi^3 u_y^* + \xi^2\frac{du_x^*}{dy} - i\xi\frac{d^2 u_y^*}{dy^2} - \frac{d^3 u_x^*}{dy^3}\right), \tag{52}$$

$$m_{yz}^*(\xi, y) = -2\mu\ell^2\left(i\xi\frac{du_y^*}{dy} + \frac{d^2 u_x^*}{dy^2}\right), \tag{53}$$

$$m_{xz}^*(\xi, y) = 2\mu\ell^2\left(-\xi^2 u_y^* + i\xi\frac{du_x^*}{dy}\right). \tag{54}$$

Now, we impose at the origin of the $(x, y)$-plane a discrete climb dislocation with Burger's vector $\mathbf{b} = (0, b, 0)$ and a discrete constrained wedge disclination with Frank's vector $\mathbf{\Omega} = (0, 0, \Omega)$. In the framework of couple-stress theory and considering the upper half-plane ($-\infty < x < \infty$, $y \geq 0$), a climb dislocation and a constrained wedge disclination give rise, respectively, to the following boundary value problems

$$u_y(x, 0^+) = -\frac{b}{2}H(x), \quad \omega(x, 0^+) = 0, \quad \sigma_{yx}(x, 0^+) = 0, \tag{55a-c}$$

$$u_y(x, 0^+) = 0, \quad \omega(x, 0^+) = \frac{\Omega}{2}H(x), \quad \sigma_{yx}(x, 0^+) = 0. \tag{56a-c}$$

where $H(x)$ is the Heaviside step-function. We emphasize once again that the term 'constrained wedge disclination' is justified from the fact that the discontinuity in rotation (cf. (56b)) does not affect the normal displacement in (56a) (see also Appendix A). Clearly, this concept departs from the one of the standard wedge disclination appearing in the settings of both classical elasticity (de Wit, 1973) and couple-stress elasticity (Anthony, 1970). This standard wedge disclination generates a field where the jump in rotation implies a discontinuity in the normal displacement too. Finally, we notice that the use of a half-plane domain (resulting from simple symmetry considerations), instead of the full-plane domain, permits the formulation of *boundary* value problems. Such a formulation provides indeed an advantage for the use of Fourier transforms.

Applying the Fourier transform to the boundary conditions (55a-c) and (56a-c), we obtain

$$u_y^*(\xi,0^+) = -b(\pi/2)^{1/2}\delta_+(\xi), \quad \omega^*(\xi,0^+) = 0, \quad \sigma_{yx}^*(\xi,0^+) = 0 . \tag{57a-c}$$

$$u_y^*(\xi,0^+) = 0, \quad \omega^*(\xi,0^+) = \Omega(\pi/2)^{1/2}\delta_+(\xi), \quad \sigma_{yx}^*(\xi,0^+) = 0 , \tag{58a-c}$$

where $\delta_+(\xi) = [\delta(\xi)/2] + [i/(2\pi\xi)]$ is the Heisenberg delta function (see e.g. Roos, 1969) and $\delta(\xi)$ is the Dirac delta distribution. However, the contribution of the Dirac delta distribution in the physical domain is only a rigid-body displacement for the problem (55) and a rigid-body rotation for the problem (56).

Next, combining (57) and (58) with (46)-(54) provides a system of algebraic equations for the functions $(A_1(\xi), A_2(\xi), A_3(\xi))$. After some algebra involving manipulations and also use of the symbolic program MATHEMATICA (version 6.0), the *transformed* displacements due to the climb dislocation and the constrained wedge disclination are found to be

$$u_x^*(\xi,y) = \frac{b}{(2\pi)^{1/2}}\left[\left(-\frac{(1-2v)}{4(1-v)|\xi|} + \frac{y}{4(1-v)} - \ell^2|\xi|\right)e^{-y|\xi|} + \ell\alpha e^{-\frac{y\alpha}{\ell}}\right]$$

$$+ \frac{i\Omega}{(2\pi)^{1/2}}\left[\frac{\ell\alpha}{\xi}e^{-\frac{y\alpha}{\ell}} - \ell^2\,\mathrm{sgn}(\xi)e^{-y|\xi|}\right], \tag{59a}$$

$$u_y^*(\xi,y) = \frac{ib}{(2\pi)^{1/2}}\left[\left(-\frac{1}{2\xi} - \frac{\mathrm{sgn}(\xi)y}{4(1-v)} + \ell^2\xi\right)e^{-y|\xi|} - \ell^2\xi e^{-\frac{y\alpha}{\ell}}\right]$$

$$+ \frac{\Omega}{(2\pi)^{1/2}}\left[\ell^2 e^{-\frac{y\alpha}{\ell}} - \ell^2 e^{-y|\xi|}\right] . \tag{59b}$$

With the aid of the inversion formula (42b) and enforcing (48)–(54), we finally obtain the expressions for the normal stress $\sigma_{yy}$ and the couple-stress $m_{yz}$ along the crack line $y = 0$ (details are given in Appendix A) which will serve as the Green's functions of the mode I crack problem, i.e.

$$\sigma_{yy}(x,y=0) = \frac{\mu b}{2\pi(1-\nu)x} + \frac{2\mu b}{\pi x}\left(\frac{2\ell^2}{x^2} - K_2\left(\frac{|x|}{\ell}\right)\right) - \frac{\mu\Omega}{\pi}\left(\frac{2\ell^2}{x^2} - K_2\left(\frac{|x|}{\ell}\right)\right) - \frac{\mu\Omega}{\pi}K_0\left(\frac{|x|}{\ell}\right),$$

(60)

$$m_{yz}(x,y=0) = -\frac{\mu b}{\pi}\left(\frac{2\ell^2}{x^2} - K_2\left(\frac{|x|}{\ell}\right)\right) - \frac{\mu b}{\pi}K_0\left(\frac{|x|}{\ell}\right) + \frac{\mu\ell\Omega}{2\pi}\mathrm{sgn}(x)\cdot G_{1,3}^{2,1}\left(\frac{x^2}{4\ell^2}\left|\begin{array}{c}1\\-1/2,\,1/2,\,0\end{array}\right.\right),$$

(61)

where $K_i(x/\ell)$ is the $i^{\text{th}}$ order modified Bessel function of the second kind and $G_{c,d}^{a,b}(\ )$ is the MeijerG function, which is a tabulated function.

Concerning now the nature of the above stress field, the following points are of notice:

(i) As $x \to 0$, the following asymptotic relations are deduced

$$\frac{2\ell^2}{x^2} - K_2\left(\frac{|x|}{\ell}\right) = O(1), \quad \frac{1}{x}\left(\frac{2\ell^2}{x^2} - K_2\left(\frac{|x|}{\ell}\right)\right) = O\left((2x)^{-1}\right), \quad K_0\left(\frac{|x|}{\ell}\right) = O(-\ln|x|),$$

$$\mathrm{sgn}(x)\cdot G_{1,3}^{2,1}\left(\frac{x^2}{4\ell^2}\left|\begin{array}{c}1\\-1/2,\,1/2,\,0\end{array}\right.\right) = O\left(-4\ell x^{-1}\right).$$

(62)

In light of the above, we conclude that as $x \to 0$, the normal stress $\sigma_{yy}$ exhibits a Cauchy-type singularity due to the climb dislocation and a logarithmic singularity due to the constrained wedge disclination. Also, as $x \to 0$, $m_{yz}$ exhibits a Cauchy singularity due to the constrained wedge disclination and a logarithmic singularity due to the climb dislocation.

(ii) As $x \to \pm\infty$, it can readily be shown that $\sigma_{yy} \to 0$ and $m_{yz} \to \mp\mu\ell\Omega$. Thus, we observe that a constrained wedge disclination does not induce normal stresses at infinity. On the contrary, the standard wedge disclination induces normal stresses that are logarithmically unbounded at infinity, in the framework of both classical elasticity (de Wit, 1973) and couple-stress elasticity (Anthony, 1970).

(iii) As $\ell \to 0$, it can be shown that the couple-stress $m_{yz}(x,y=0)$ vanishes, while the normal stress $\sigma_{yy}(x,y=0)$ degenerates into the field $\mu b/2\pi(1-\nu)x$ (first term in the RHS of equation (60)) given by a classical elasticity analysis for a discrete climb dislocation. Thus, we see that a constrained wedge disclination induces stresses and couple-stresses *only* when the material length is $\ell \neq 0$, i.e. when couple-stress effects are taken into account. This is a convenient feature of the Green's functions in (60) and (61) since, in the limit $\ell \to 0$, the respective Green's function of classical elasticity (i.e. the field induced by a discrete climb dislocation) is recovered.

## 6. Reduction of the crack problem to a system of singular integral equations: Results

The corrective stresses (41a-c) are generated by a continuous distribution of climb dislocations and constrained wedge disclinations along the faces of the crack. The normal stress $\sigma_{yy}$ and the couple-stress $m_{yz}$ induced by a continuous distribution of dislocations and disclinations can be derived by integrating the field (along the crack-faces) of a discrete climb dislocation (Eq. (60)) and a discrete constrained wedge disclination (Eq. (61)). We note that (41c) is automatically satisfied since neither the discrete dislocation nor the discrete disclination produce shear stresses $\sigma_{yx}$ along the crack-line $y = 0$. Then, satisfaction of the boundary conditions (41a) and (41b) results in a system of coupled integral equations, which govern the problem. Separating the singular from the regular parts of the kernels, we finally obtain the following system of singular integral equations

$$-\sigma_0 = \frac{\mu(3-2\nu)}{2\pi(1-\nu)} \int_{-a}^{a} \frac{B(\xi)}{x-\xi} d\xi + \frac{\mu}{\pi a} \int_{-a}^{a} W(\xi) \cdot \ln\frac{|x-\xi|}{\ell} d\xi + \frac{2\mu}{\pi a} \int_{-a}^{a} B(\xi) \cdot k_1(x,\xi) d\xi$$

$$- \frac{\mu}{\pi a} \int_{-a}^{a} W(\xi) \cdot k_2(x,\xi) d\xi , \qquad |x| < a , \qquad (63a)$$

$$0 = -\frac{2\mu\ell^2}{\pi a} \int_{-a}^{a} \frac{W(\xi)}{x-\xi} d\xi + \frac{\mu}{\pi} \int_{-a}^{a} B(\xi) \cdot \ln\frac{|x-\xi|}{\ell} d\xi - \frac{\mu}{\pi} \int_{-a}^{a} B(\xi) \cdot k_2(x,\xi) d\xi$$

$$+ \frac{\mu\ell}{2\pi a} \int_{-a}^{a} W(\xi) \cdot k_3(x,\xi) d\xi , \qquad |x| < a , \qquad (63b)$$

where $B(\xi)$ and $W(\xi)$ are, respectively, the dislocation and disclination densities defined as

$$B(\xi) = \frac{db(\xi)}{d\xi} = -\frac{d\Delta u_y(\xi)}{d\xi} , \qquad \Delta u_y(x) = -\int_{-a}^{x} B(\xi) d\xi , \qquad (64a)$$

$$W(\xi) = a\frac{d\Omega(\xi)}{d\xi} = a\frac{d\Delta\omega(\xi)}{d\xi} , \qquad \Delta\omega(x) = \frac{1}{a}\int_{-a}^{x} W(\xi) d\xi , \qquad (64b)$$

and the kernels $k_\beta(x,\xi)$, with $\beta = 1,2,3$, are defined as

$$k_1(x,\xi) = \frac{a}{x-\xi}\left[\frac{2\ell^2}{(x-\xi)^2} - K_2(|x-\xi|/\ell) - \frac{1}{2}\right], \tag{65a}$$

$$k_2(x,\xi) = \left[\frac{2\ell^2}{(x-\xi)^2} - K_2(|x-\xi|/\ell)\right] + \left[K_0(|x-\xi|/\ell) + \ln(|x-\xi|/\ell)\right], \tag{65b}$$

$$k_3(x,\xi) = \mathrm{sgn}(x-\xi)\cdot G_{1,3}^{2,1}\!\left(\left.\frac{(x-\xi)^2}{4\ell^2}\right|\begin{matrix}1\\-1/2,\,1/2,\,0\end{matrix}\right) + \frac{4\ell}{x-\xi}. \tag{65c}$$

In the above relations, $\Delta u_y(x)$ represents the relative opening displacement and $\Delta\omega(x)$ the relative rotation between the upper and lower crack faces. Furthermore, it is noted that both densities are dimensionless according to (64).

Also, using the asymptotic expansions of the modified Bessel functions (see e.g. Erdelyi 1953), it can readily be shown that the first two kernels (Eqs. (65a,b) are regular as $x \to \xi$ and $\ell > 0$. To understand now the nature of the third kernel (Eq. (65c)), we expand the MeijerG function, with the aid of the symbolic program MATHEMATICA (version 6.0), in series as $x \to \xi$, and have

$$\mathrm{sgn}(x-\xi)\cdot G_{1,3}^{2,1}\!\left(\left.\frac{(x-\xi)^2}{4\ell^2}\right|\begin{matrix}1\\-1/2,\,1/2,\,0\end{matrix}\right) = -\frac{4\ell}{x-\xi} + (a_1 + a_2\ln|x-\xi|)\cdot(x-\xi) + O((x-\xi)^3 \ln|x-\xi|),$$

$$(66)$$

where $(a_1, a_2)$ are constants depending on the characteristic material length $\ell$. Since $\lim_{x\to\xi}(x-\xi)^n \cdot \ln|x-\xi| = 0$ for $n > 0$, it is apparent that the kernel $k_3(x,\xi)$ is a regular kernel ($\ell > 0$) in the closed interval $-a \leq (x,\xi) \leq a$.

As is standard in the DDT (see e.g. Hills et al., 1996), the unknown densities $B(\xi)$ and $W(\xi)$ can be written as a product of a regular bounded function and a singular function characterizing the asymptotic behavior near the crack tips. Within the framework of couple-stress elasticity, asymptotic analysis near a mode I crack tip (Huang et al., 1997) showed that both the crack-face displacement $u_y$ and the rotation $\omega$ behave as $r^{1/2}$ in the crack tip region, where $r$ is the polar distance from the crack tip. Such a behavior was also corroborated by the uniqueness theorem for crack problems of couple-stress elasticity which imposes the requirement of boundedness for both crack-tip displacement and rotation (Grentzelou and Georgiadis, 2005). Accordingly, the dislocation and the disclination densities are expressed in the following form

$$B(\xi) = f(\xi)/(a^2 - \xi^2)^{1/2}, \quad W(\xi) = g(\xi)/(a^2 - \xi^2)^{1/2},  \tag{67}$$

where $f(\xi)$ and $g(\xi)$ are regular bounded functions in the interval $|\xi| \leq a$. Further, in order to ensure uniqueness of the values of the normal displacement and the rotation for a closed loop around the crack, the following closure conditions must be satisfied (the first of them is standard in the DDT applied to classical elasticity)

$$\int_{-a}^{a} B(x)\,dx = 0, \quad \int_{-a}^{a} W(x)\,dx = 0. \tag{68a,b}$$

Before proceeding to the numerical solution of the system (63), it is interesting to consider two limit cases concerning the behavior of this system w.r.t. limit values of the characteristic length $\ell$.

First, by letting $\ell \to 0$, it can readily be shown that the integral equation in (63b) vanishes identically, whereas the one in (63a) degenerates into the counterpart equation governing the mode I crack problem of classical elasticity. The latter equation is as follows

$$-\sigma_0 = \frac{\mu}{2\pi(1-\nu)} \int_{-a}^{a} \frac{B(\xi)}{x - \xi}\,d\xi, \quad |x| < a. \tag{69}$$

Secondly, we let $\ell \to \infty$. Then, by multiplying (63b) with $(1/\ell^2)$ and noting that

$$\lim_{\ell \to \infty} \frac{1}{\ell^2} \ln \frac{|x-\xi|}{\ell} = 0, \quad \lim_{\ell \to \infty} \frac{1}{\ell^2} k_2(x,\xi) = 0, \quad \lim_{\ell \to \infty} \frac{1}{\ell^2} k_3(x,\xi) = 0, \tag{70}$$

we find that the integral equation in (63b) takes the following form

$$\int_{-a}^{a} \frac{W(\xi)}{x - \xi}\,d\xi = 0, \quad |x| < a, \tag{71}$$

which along with (67b) and the closure condition (68b) has the unique solution $W(\xi) \equiv 0$. Now, in light of the above and noting also that $\lim_{\ell \to \infty} k_1(x,\xi) = 0$, the system (63) degenerates as $\ell \to \infty$ to the following *single* singular integral equation

$$-\sigma_0 = \frac{\mu(3-2v)}{2\pi(1-v)} \int_{-a}^{a} \frac{B(\xi)}{x-\xi} d\xi, \qquad |x| < a, \tag{72}$$

Further, it can be readily be shown, that the ratio of the crack-face displacements obtained by the solutions of, respectively, (72) ($\ell \to \infty$ case) and (69) ($\ell \to 0$ case) is $1/(3-2v)$. The same ratio was also obtained by Sternberg and Muki (1967) for the mode I problem and by Gourgiotis and Georgiadis (2007) for the mode II problem in couple-stress elasticity. Of course, from the physical point of view, the case $\ell \to \infty$ is of no interest since the characteristic length is a small quantity. Nonetheless, the latter result for the ratio of displacements shows mathematically that there is a lower bound for the crack-face displacement when $\ell \to \infty$.

For the numerical solution of the system of singular integral equations in (63), the Gauss-Chebyshev quadrature proposed by Erdogan and Gupta (1972) is employed, with a modification that takes into account the logarithmic kernel (details are given in Appendix B). In particular, after the appropriate normalization over the interval $[-1, 1]$, this system takes the following discretized form

$$-\sigma_0 = \frac{\mu(3-2v)}{2(1-v)n} \sum_{i=1}^{n} \frac{f(s_i)}{t_k - s_i} + \frac{\mu}{n} \sum_{i=1}^{n} g(s_i) \cdot \ln(p|t_k - s_i|) + \frac{2\mu}{n} \sum_{i=1}^{n} f(s_i) \cdot k_1(at_k, as_i)$$

$$- \frac{\mu}{n} \sum_{i=1}^{n} g(s_i) \cdot k_2(at_k, as_i) + \frac{\mu}{\pi} G_n(t_k) \cdot T_n(t_k) \sum_{i=1}^{n} \frac{g(s_i)}{(t_k - s_i)T'_n(s_i)}, \tag{73a}$$

$$0 = -\frac{2\mu}{p^2 n} \sum_{i=1}^{n} \frac{g(s_i)}{t_k - s_i} + \frac{\mu}{n} \sum_{i=1}^{n} f(s_i) \cdot \ln(p|t_k - s_i|) - \frac{\mu}{n} \sum_{i=1}^{n} f(s_i) \cdot k_2(at_k, as_i)$$

$$+ \frac{\mu}{2pn} \sum_{i=1}^{n} g(s_i) \cdot k_3(at_k, as_i) + \frac{\mu}{\pi} G_n(t_k) \cdot T_n(t_k) \sum_{i=1}^{n} \frac{f(s_i)}{(t_k - s_i)T'_n(s_i)}, \tag{73b}$$

where $p = a/\ell$, $t = x/a$, $s = \xi/a$. The integration and collocation points are given, respectively, as

$$T_n(s_i) = 0, \qquad s_i = \cos[(2i-1)\pi/2n], \qquad i = 1,\ldots,n, \tag{74a}$$

$$U_{n-1}(t_k) = 0, \qquad t_k = \cos[k\pi/n], \qquad k = 1,\ldots,n-1, \tag{74b}$$

where $T_n(x)$ and $U_n(x)$ are the Chebyshev polynomials of the first and second kind, respectively. The function $G_n(x)$ in the last term of (73a) and (73b) is the quadrature error due to the existence of the logarithmic kernel and is defined in Appendix B. In fact, introducing this function greatly improves the speed of convergence of the solution of the above system. Now, (73a) and (73b) together with the auxiliary conditions (68) provide an algebraic system of $2n$ equations in the $2n$ unknown functions $f(s_i)$ and $g(s_i)$. A computer program was written that solved this system.

Now, some numerical results will be presented. Figure 2 depicts the influence of the ratio $a/\ell$ on the normal crack-face displacement. It is noteworthy that as the crack length becomes comparable to the characteristic length $\ell$, the material exhibits a more *stiff* behavior, i.e. the crack-face displacement becomes smaller in magnitude. We note further that the displacements obtained from the classical elasticity solution are an *upper bound* for those obtained from the present couple-stress elasticity solution.

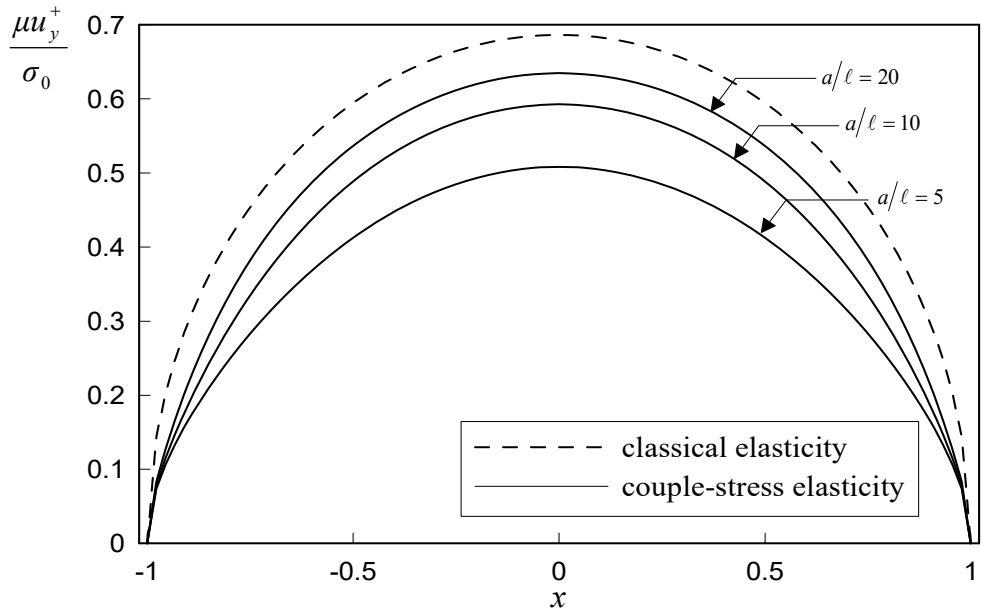

**Fig. 2** Normalized upper-half normal crack displacement profile. The Poisson's ratio is $v = 0.3$.

Figure 3 depicts the influence of the ratio $a/\ell$ on the crack-face rotation. We note that as $\ell \to 0$ the rotation in the crack-tip vicinity tends to the unbounded limit of classical elasticity. This indicates a typical *boundary layer* behavior in the couple-stress solution.

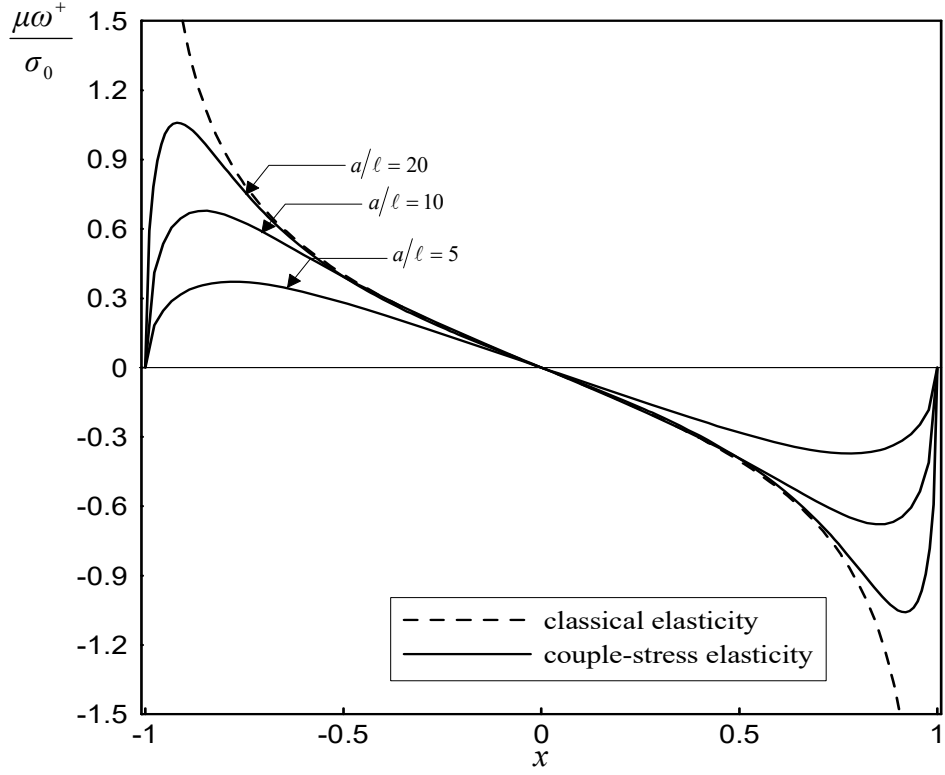

**Fig. 3** Normalized upper-half crack rotation. The Poisson's ratio is $v = 0.3$.

Next, the behavior of the normal stress as given by (63a) will be determined. We have

$$\sigma_{yy}(|x|>a, y=0) = \sigma_0 + \frac{\mu(3-2v)}{2\pi(1-v)} \int_{-a}^{a} \frac{B(\xi)}{x-\xi} d\xi + \frac{\mu}{\pi a} \int_{-a}^{a} W(\xi) \cdot \ln\frac{|x-\xi|}{\ell} d\xi + \frac{2\mu}{\pi a} \int_{-a}^{a} B(\xi) \cdot k_1(x,\xi) d\xi$$

$$- \frac{\mu}{\pi a} \int_{-a}^{a} W(\xi) \cdot k_2(x,\xi) d\xi . \qquad (75)$$

Due to the symmetry of the problem (in geometry and loading) with respect to $y$-axis, we confine attention only to the right crack tip. As $x \to a^+$, the following asymptotic relations hold

$$\int_{-a}^{a} W(\xi) \cdot \ln|x-\xi| d\xi = O(1) , \quad \int_{-a}^{a} B(\xi) \cdot k_1(x,\xi) d\xi = O(1) , \quad \int_{-a}^{a} W(\xi) \cdot k_2(x,\xi) d\xi = O(1) ,$$

$$\int_{-a}^{a} \frac{B(\xi)}{x-\xi} d\xi = O\left((x-a)^{-1/2}\right), \qquad (x > a) , \tag{76}$$

where the dislocation and the disclination densities are defined in (67). In view of the above, we conclude that $\sigma_{yy}$ exhibits a square root singularity at the crack tips just as in classical elasticity. Figure 4 now depicts the distribution of the normal stress ahead of the RHS crack tip. Normalized quantities are used and $K_I^{clas.}$ denotes the stress intensity factor provided by the classical elasticity solution. For convenience, a new variable $\bar{x} = x - a$ is introduced measuring distance from the RHS crack tip. We observe that the couple-stress effects are dominant within a zone of length $2\ell$, whereas outside this zone $\sigma_{yy}$ gradually approaches the distribution given by the classical solution. It is also noted that the normal stress $\sigma_{yy}$ in (75) depends not only upon the ratio $a/\ell$ but also upon the Poisson's ratio $v$. This was also observed by Sternberg and Muki (1967).

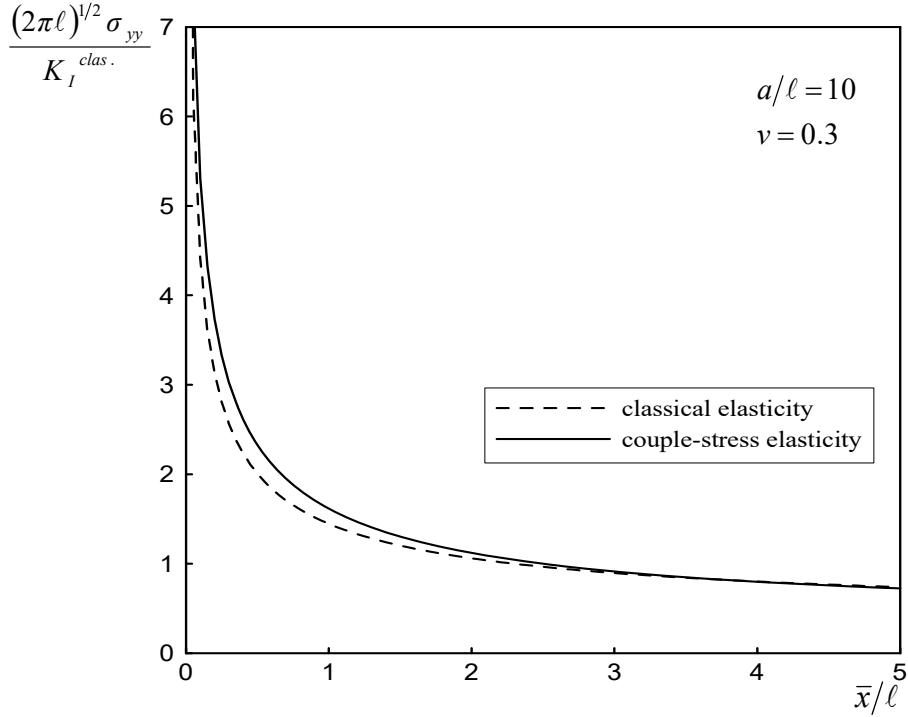

**Fig. 4** Distribution of the normal stress ahead of the crack tip for $a/\ell = 10$ and $v = 0.3$.

Figure 5 depicts the variation of the ratio $K_I/K_I^{clas.}$ with $\ell/a$ for three different values of the Poisson's ratio. The stress intensity factor in couple-stress elasticity is defined as $K_I = \lim_{x \to a^+}[2\pi(x-a)]^{1/2}\sigma_{yy}(x,0)$ with $\sigma_{yy}(x,0)$ being given by (75). It is observed that for a material with $a/\ell = 20$ and Poisson's ratio $\nu = 0.5$, there is a 18% increase in the stress intensity factor when couple-stress effects are taken into account, while for $\nu = 0.25$ and $\nu = 0$ the increase becomes 24,3% and 29,5%, respectively. It should be noted that when $\ell/a = 0$ (no couple-stress effects) the above ratio should evidently become $K_I/K_I^{clas.} = 1$. Therefore, the stress-ratio plotted in Fig. 5 exhibits a finite jump *discontinuity* at the limit $\ell/a = 0$; the ratio at the tip of the crack rises abruptly as $\ell/a$ departs from zero. The same discontinuity was observed by Sternberg and Muki (1967), who attributed this behavior to the severe boundary layer effects of couple-stress elasticity in singular stress-concentration problems. Finally, it is noted that the ratio decreases monotonically with increasing values of $\ell/a$ and tends to unity as $\ell/a \to \infty$.

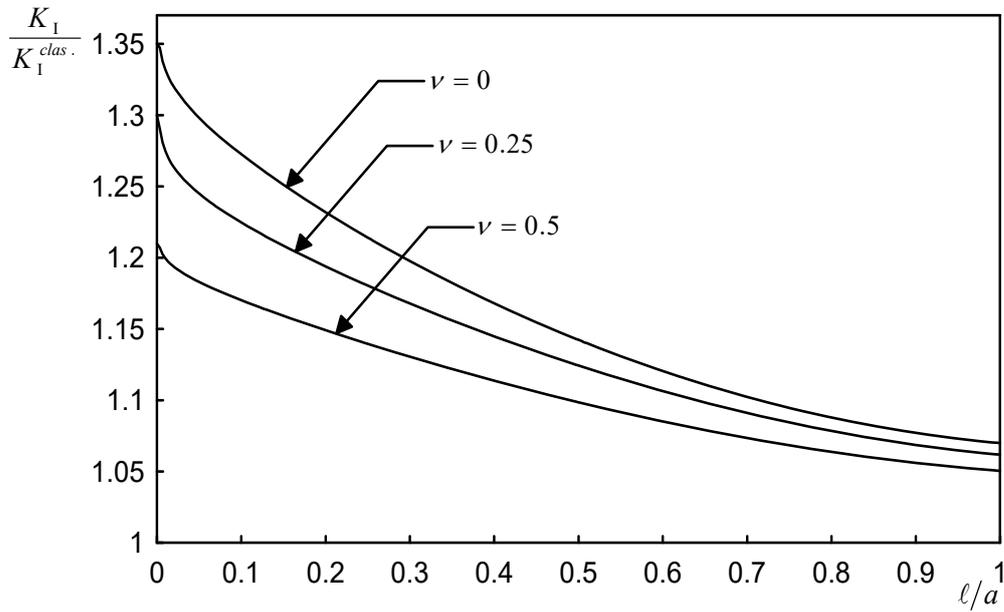

**Fig. 5** Variation of the ratio of stress intensity factors in couple-stress elasticity and classical elasticity with $\ell/a$.

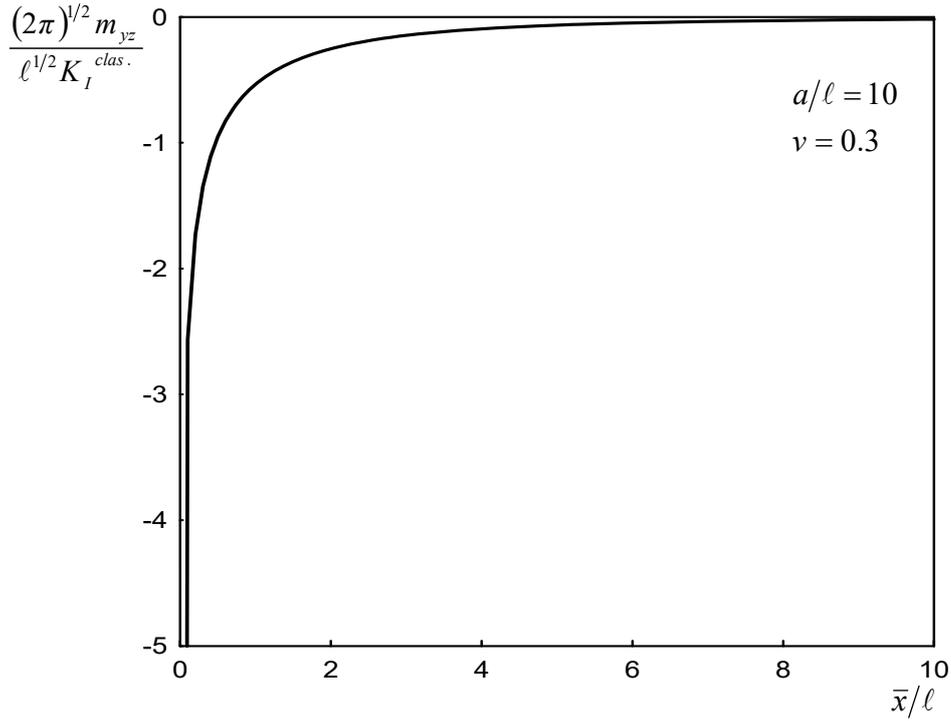

**Fig. 6** Distribution of couple-stress ahead of the crack tip for $a/\ell = 10$ and $v = 0.3$.

The behavior of the couple-stress $m_{yz}$ will be examined next. From the previous analysis, we have

$$m_{yz}(|x|>a, y=0) = -\frac{2\mu\ell^2}{\pi a}\int_{-a}^{a}\frac{W(\xi)}{x-\xi}d\xi + \frac{\mu}{\pi}\int_{-a}^{a}B(\xi)\cdot\ln\frac{|x-\xi|}{\ell}d\xi - \frac{\mu}{\pi}\int_{-a}^{a}B(\xi)\cdot k_2(x,\xi)d\xi$$

$$+ \frac{\mu\ell}{2\pi a}\int_{-a}^{a}W(\xi)\cdot k_3(x,\xi)d\xi \ . \tag{77}$$

Focusing attention again to the RHS crack tip, the following asymptotic relations for $x \to a^+$ were found to hold ($x > a$)

$$\int_{-a}^{a}B(\xi)\cdot\ln|x-\xi|d\xi = O(1) \ , \quad \int_{-a}^{a}B(\xi)\cdot k_2(x,\xi)d\xi = O(1) \ , \quad \int_{-a}^{a}W(\xi)\cdot k_3(x,\xi)d\xi = O(1) \ ,$$

$$\int_{-a}^{a}\frac{W(\xi)}{x-\xi}d\xi = O\left((x-a)^{-1/2}\right) \ , \tag{78a-d}$$

which leads us to the conclusion that the couple-stress $m_{yz}$ behaves like $\sim \bar{x}^{-1/2}$ in the vicinity of the crack-tip (the variable $\bar{x} = x - a$ measures distance from the RHS crack tip). This is in agreement with the asymptotic results of Huang et al. (1997).

Figure 6 depicts (with the use of normalized quantities) the distribution of the couple-stress ahead of the RHS crack tip. It should further be noted that the stresses and couple-stresses at any point of the cracked body can be evaluated through integration along the crack-faces of Eqs. (A7)-(A12) (see Appendix A), once the dislocation and disclination densities are known. The latter equations are the *full-field* Green's functions for the mode I crack problem in couple-stress elasticity.

## 7. Evaluation of the *J*- integral

In this Section, we evaluate the $J$-integral (energy release rate) of Fracture Mechanics and examine its dependence upon the ratio of lengths $\ell/a$ and the Poisson's ratio $\nu$. The path-independent $J$-integral within couple-stress elasticity was first established by Atkinson and Leppington (1974) (see also Atkinson and Leppington, 1977; Lubarda and Markenskoff, 2000) and is written as

$$J = \int_\Gamma \left[ Wn_x - T_q \frac{\partial u_q}{\partial x} - M_q \frac{\partial \omega_q}{\partial x} \right] d\Gamma = \int_\Gamma \left[ Wn_x - P_q \frac{\partial u_q}{\partial x} - R_q \frac{\partial \omega_q}{\partial x} \right] d\Gamma$$

$$= \int_\Gamma \left( Wdy - \left[ P_q \frac{\partial u_q}{\partial x} + R_q \frac{\partial \omega_q}{\partial x} \right] d\Gamma \right), \qquad (79)$$

where $\Gamma$ is a piece-wise smooth simple two-dimensional contour surrounding the crack-tip, $W$ is the strain-energy density, $u_q$ is the displacement vector, $\omega_q$ is the rotation vector, $(T_q, M_q)$ are the tractions defined in (3) and (4), and $(P_q, R_q)$ are the reduced force-traction and the tangential couple-traction defined in (18) and (19).

For the evaluation of the $J$-integral, we consider the rectangular-shaped contour $\Gamma$ (surrounding the RHS crack-tip) with vanishing 'height' along the $y$- direction and with $\varepsilon \to +0$ (see Fig. 7). Such a contour was first introduced by Freund (1972) in examining the energy flux into the tip of a rapidly extending crack and it was proved particularly convenient in computing energy quantities in the vicinity of crack tips (see e.g. Burridge, 1976; Georgiadis, 2003). In fact, this type of contour permits using solely the *asymptotic* near-tip stress and displacement fields. It is noted that upon this choice of contour, the integral $\int_\Gamma W\, dy$ in (79) becomes zero if we allow the 'height' of the rectangle to vanish. In this way, the expression for the $J$-integral becomes

$$J = -2\lim_{\varepsilon \to +0}\left\{\int_{a-\varepsilon}^{a+\varepsilon}\left(P_q\frac{\partial u_q}{\partial x} + R_q\frac{\partial \omega_q}{\partial x}\right)dx\right\}. \tag{80}$$

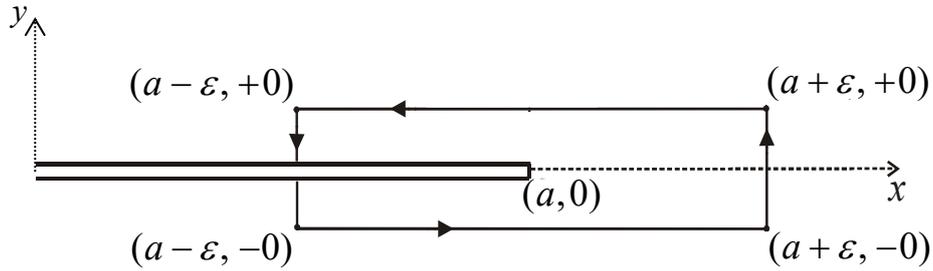

**Fig. 7** Rectangular-shaped contour surrounding the RHS crack-tip.

Further, we take into account that in the mode I case the shear stress $\sigma_{yx}$ is zero along the crack line $(y=0)$ and the crack-faces are defined by $\mathbf{n}=(0,\pm 1)$. Then, the $J$-integral gets the following form

$$J = -2\lim_{\varepsilon \to +0}\left\{\int_{a-\varepsilon}^{a+\varepsilon}\left(\sigma_{yy}(x,y=0^+)\cdot\frac{\partial u_y(x,y=0^+)}{\partial x} + m_{yz}(x,y=0^+)\cdot\frac{\partial \omega(x,y=0^+)}{\partial x}\right)dx\right\}. \tag{81}$$

Now, the dominant singular behavior (in the vicinity of the crack-tip) of the normal stress $\sigma_{yy}$ and the couple-stress $m_{yz}$ is due to the Cauchy integrals in (75) and (77), respectively. These stresses are written as

$$\sigma_{yy}(x\to a^+, y=0^+) = \lim_{x\to a^+}\frac{\mu(3-2\nu)}{2\pi(1-\nu)}\int_{-a}^{a}\frac{B(\xi)}{x-\xi}d\xi = \lim_{t\to 1^+}\frac{\mu(3-2\nu)}{2\pi(1-\nu)}\int_{-1}^{1}\frac{f(s)}{(1-s^2)^{1/2}(t-s)}ds$$

$$= \frac{\mu(3-2\nu)}{2(1-\nu)}\frac{f(1)}{2^{1/2}}\cdot(t-1)^{-1/2}, \quad (t>1), \tag{82}$$

$$m_{yz}(x \to a^+, y = 0^+) = -\lim_{x \to a^+} \frac{2\mu \ell^2}{\pi a} \int_{-a}^{a} \frac{W(\xi)}{x - \xi} d\xi = -\lim_{t \to 1^+} \frac{2\mu \ell^2}{\pi a} \int_{-1}^{1} \frac{g(s)}{(1-s^2)^{1/2}(t-s)} ds$$

$$= -\frac{2\mu \ell^2}{a} \frac{g(1)}{2^{1/2}} \cdot (t-1)^{-1/2}, \quad (t > 1). \tag{83}$$

The regular functions $f(s)$ and $g(s)$ were defined in (67) and their values at the crack-tips ($t = \pm 1$) can be evaluated by the use of Krenk's interpolation technique (Krenk, 1975). Also, the limits of the integrals in (82) and (83) are obtained by the use of the following asymptotic relation (see e.g. Muskheleshvili, 1958)

$$\lim_{t \to 1^+} \int_{-1}^{1} \frac{h(s)}{(1-s^2)^{1/2}(t-s)} ds = \frac{\pi}{2^{1/2}} h(1)(t-1)^{-1/2}, \quad (t > 1), \tag{84}$$

where $h(s)$ is a regular bounded function in the interval $|s| \leq 1$.

Also, in view of the definitions in (64), the following asymptotic relations are established

$$\frac{\partial u_y(x \to a^-, y = 0^+)}{\partial x} = -\frac{1}{2} \lim_{x \to a^-} B(x) = -\frac{1}{2} \frac{f(1)}{2^{1/2}} (1-t)^{-1/2}, \quad (t < 1), \tag{85}$$

$$\frac{\partial \omega(x \to a^-, y = 0^+)}{\partial x} = \frac{1}{2a} \lim_{x \to a^-} W(x) = \frac{1}{2a} \frac{g(1)}{2^{1/2}} (1-t)^{-1/2}, \quad (t < 1), \tag{86}$$

Then, the above results allow us to write the $J$-integral as[1]

$$J = -2a \lim_{\varepsilon \to 0} \left\{ -\frac{\mu(3-2\nu)}{8(1-\nu)} f^2(1) \cdot \int_{-\varepsilon/a}^{\varepsilon/a} (t_+)^{-1/2} \cdot (t_-)^{-1/2} d\bar{t} - \frac{\mu}{2} \left(\frac{\ell}{a}\right)^2 g^2(1) \cdot \int_{-\varepsilon/a}^{\varepsilon/a} (t_+)^{-1/2} \cdot (t_-)^{-1/2} d\bar{t} \right\}$$

$$= \frac{\mu \pi}{2} a \left\{ \frac{(3-2\nu)}{4(1-\nu)} f^2(1) + \left(\frac{\ell}{a}\right)^2 g^2(1) \right\}, \tag{87}$$

---

[1] Note that in Gourgiotis, P.A., Georgiadis, H.G., 2008. An approach based on distributed dislocations and disclinations for crack problems in couple-stress elasticity. *Int. J. Solids Struct.* 45, 5521-5539, Eq. (87) contained a typographical error i.e. the term (1-$\nu$) in the denominator was inadvertently omitted.

where $\bar{t} = t - 1$ and, for any real $\lambda$ with the exception of $\lambda = -1, -2, -3, \ldots$, the following definitions of the distributions (of the bisection type) $t_+^\lambda$ and $t_-^\lambda$ are employed (see e.g. Gelfand and Shilov, 1964)

$$t_+^\lambda = \begin{cases} |\bar{t}|^\lambda, & \text{for } \bar{t} > 0 \\ 0, & \text{for } \bar{t} < 0 \end{cases} \quad \text{and} \quad t_-^\lambda = \begin{cases} 0, & \text{for } \bar{t} > 0 \\ |\bar{t}|^\lambda, & \text{for } \bar{t} < 0 \end{cases}. \quad (88a,b)$$

It is further noted that the product of distributions inside the integrals in (87) is obtained here by the use of Fisher's theorem (Fisher, 1971), i.e. the operational relation $(t_+)^{-1-\lambda}(t_-)^\lambda = -\pi \delta(\bar{t})[2\sin(\pi\lambda)]^{-1}$ with $\lambda \neq -1, -2, -3, \ldots$ and $\delta(t)$ being the Dirac delta distribution. Use is also made of the fundamental property of the Dirac delta distribution that $\int_{-\varepsilon}^{\varepsilon} \delta(t) dt = 1$.

From the above analysis, we were able to evaluate the $J$-integral. Our results are shown graphically in Figure 8. The graph depicts the dependence of the ratio $J/J^{clas.}$ upon the ratio of lengths $\ell/a$ for three different values of the Poisson's ratio of the material. $J^{clas.} \equiv \pi(1-v^2)\sigma_0^2 a/E$ is the respective integral in classical elasticity (see e.g. Rice, 1968). The calculations show that as $\ell/a \to 0$, the $J$-integral in couple-stress elasticity tends continuously to its counterpart in classical elasticity. This behavior was previously observed by Atkinson and Leppington (1977), who followed a different analysis than the present one. Also, $J < J^{clas.}$ for $\ell \neq 0$. The latter result seems to be a consequence of the fact that the crack-face displacements and rotations (see Figs. 2 and 3) are significantly smaller than the respective ones in classical theory. This not only compensates the increase of the normal stress ahead of the crack-tip (this stress aggravation in couple-stress elasticity is shown in Fig. 4), but it results evidently in an overall decrease of the energy release rate when couple-stress effects are taken into account. We also found that $J/J^{clas.}$ decreases monotonically with increasing values of $\ell/a$ and tends to the limit $1/(3-2v)$ as $\ell/a \to \infty$.

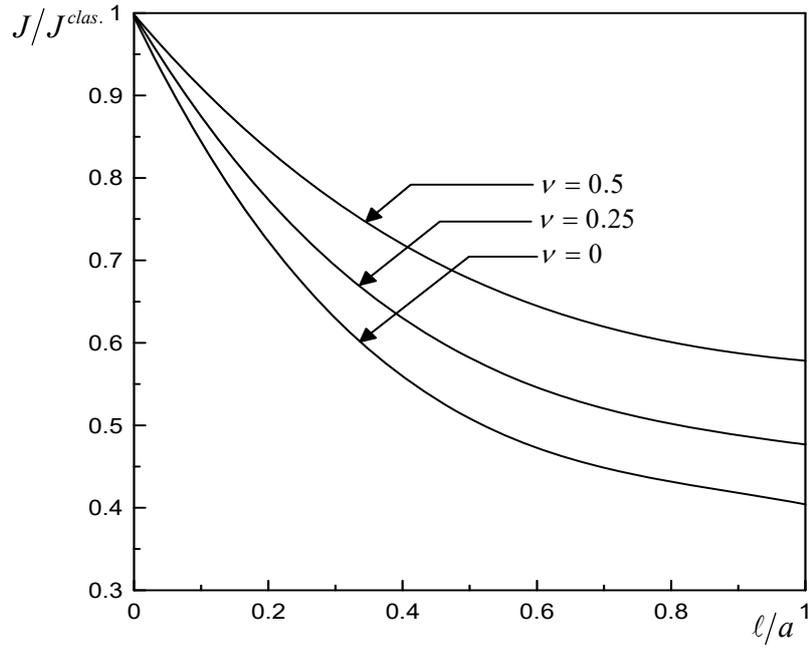

**Fig. 8** Variation of the ratio of the $J$-integral in couple-stress elasticity and in classical elasticity with $\ell/a$.

## 8. Concluding remarks

In this paper, the technique of distributed dislocations was extended in couple-stress elasticity for the solution of the mode I crack problem. Contrary to classical elasticity where a distribution of climb dislocations suffices to model the mode I crack problem, here (due to the nature of the boundary conditions that arise in couple-stress elasticity) introducing an additional discontinuity (the so-called constrained wedge disclination) was necessary to solve the problem. Considering a continuous distribution of climb dislocations and constrained wedge disclinations along the crack faces results in a coupled system of singular integral equations with both Cauchy-type and logarithmic kernels. This system of equations was solved numerically and a full-field solution was obtained.

The proposed technique provides for crack problems an efficient alternative to the elaborate analytical method of dual integral equations and the numerical methods of Finite and Boundary Elements. Especially with the latter two methods, one may encounter difficulties when dealing with crack problems in generalized continua. Also, the proposed technique is appropriate for problems with finite-length cracks where the standard Wiener-Hopf technique meets with serious difficulties (the Wiener-Hopf technique suits better problems with semi-infinite cracks). The present approach has the advantage that it provides results not only restricted to the near-tip region – it may give full-field results.

The present results indicate that the material microstructure of the couple-stress type has generally rigidity (smaller crack-face displacements and rotations) and strengthening (stress aggravation ahead of the crack-tip) effects. In particular, the crack-face displacement becomes significantly smaller than that in classical elasticity, when the crack length $2a$ is comparable to the characteristic length $\ell$ of the material (it decreases about 30% for $a/\ell = 5$). Also, it is observed that the stress intensity factor $K_I$ is higher than the one predicted by classical elasticity. In particular, for a material with $a/\ell = 20$ and Poisson's ratio $\nu = 0.25$ there is a 24.3% increase when couple-stress effects are taken into account, whereas for $\nu = 0$ and $\nu = 0.5$ the increase is 29.5% and 18%, respectively. Finally, the $J$-integral in couple-stress elasticity tends continuously to its counterpart in classical elasticity as $\ell/a \to 0$. For $\ell \neq 0$, a decrease of its value is noticed in comparison with the classical theory and this indicates that the rigidity effect dominates over the strengthening effect in the energy release rate. The $J$-integral decreases monotonically with increasing values of $\ell/a$ and tends to a certain limit as $\ell/a \to \infty$.

**Appendix A: Displacements and stresses for a climb dislocation and a constrained wedge disclination**

In this Appendix, we derive the displacement, rotation, stress and couple-stress fields due to a discrete climb dislocation and a discrete constrained wedge disclination situated at the origin of a full space in a material governed by plane-strain couple-stress elasticity. The fields apply for *any* point (not only along the line $y = 0$) of the full space.

Using the Fourier inversion formula in (42b), we obtain from (59) the following integral representation of the displacement field for a climb dislocation and a constrained wedge disclination

$$u_x(x,y) = \frac{b}{4\pi(1-\nu)} \int_0^\infty \left(-\frac{(1-2\nu)}{\xi} + y\right) e^{-y\xi} \cos(\xi x) d\xi - \frac{b\ell^2}{\pi} \int_0^\infty \xi e^{-y\xi} \cos(\xi x) d\xi$$

$$+ \frac{b\ell}{\pi} \int_0^\infty (1+\ell^2\xi^2)^{1/2} e^{-\frac{y(1+\ell^2\xi^2)^{1/2}}{\ell}} \cos(\xi x) d\xi - \frac{\Omega\ell^2}{\pi} \int_0^\infty e^{-y\xi} \sin(\xi x) d\xi$$

$$+\frac{\Omega \ell}{\pi}\int_0^\infty \frac{(1+\ell^2\xi^2)^{1/2}}{\xi} e^{-\frac{y(1+\ell^2\xi^2)^{1/2}}{\ell}} \sin(\xi x)d\xi \ , \tag{A1}$$

$$u_y(x,y) = -\frac{b}{\pi}\int_0^\infty \left(\frac{1}{2\xi}+\frac{y}{4(1-v)}\right)e^{-y\xi}\sin(\xi x)d\xi + \frac{b\ell^2}{\pi}\int_0^\infty \xi e^{-y\xi}\sin(\xi x)d\xi$$

$$-\frac{b\ell^2}{\pi}\int_0^\infty \xi e^{-\frac{y(1+\ell^2\xi^2)^{1/2}}{\ell}}\sin(\xi x)d\xi - \frac{\Omega \ell^2}{\pi}\int_0^\infty e^{-y\xi}\cos(\xi x)d\xi$$

$$+\frac{\Omega \ell^2}{\pi}\int_0^\infty e^{-\frac{y(1+\ell^2\xi^2)^{1/2}}{\ell}}\cos(\xi x)d\xi \ . \tag{A2}$$

The above integrals are computed by invoking results from the theory of distributions (see e.g. Zemanian, 1965; Roos, 1969). In particular, we have

$$I_1 \equiv \int_0^\infty \frac{e^{-y\xi}}{\xi}\sin(\xi x)d\xi = \tan^{-1}(x/y) \ , \quad I_2 \equiv \int_0^\infty e^{-y\xi}\cos(\xi x)d\xi = \frac{y}{r^2} \ ,$$

$$I_3 \equiv \int_0^\infty e^{-y\xi}\sin(\xi x)d\xi = \frac{x}{r^2} \ , \quad I_4 \equiv \int_0^\infty \xi e^{-y\xi}\cos(\xi x)d\xi = \frac{y^2-x^2}{r^4} \ ,$$

$$I_5 \equiv \int_0^\infty \xi e^{-y\xi}\sin(\xi x)d\xi = \frac{2xy}{r^4} \ , \quad I_6 \equiv \int_0^\infty e^{-\frac{y(1+\ell^2\xi^2)^{1/2}}{\ell}}\cos(\xi x)d\xi = \frac{yK_1(r/\ell)}{r\ell} \ ,$$

$$I_7 \equiv \int_0^\infty \xi e^{-\frac{y(1+\ell^2\xi^2)^{1/2}}{\ell}}\sin(\xi x)d\xi = -\frac{\partial}{\partial x}(I_6) = \frac{xy}{r^2\ell^2}K_2(r/\ell) \ ,$$

$$I_8 \equiv \int_0^\infty (1+\ell^2\xi^2)^{1/2} e^{-\frac{y(1+\ell^2\xi^2)^{1/2}}{\ell}}\cos(\xi x)d\xi = -\ell\frac{\partial}{\partial y}(I_6) = \frac{y^2 K_2(r/\ell)}{r^2\ell} - \frac{K_1(r/\ell)}{r} \ ,$$

$$I_9 \equiv \int_0^\infty \frac{e^{-y\xi}}{\xi}\cos(\xi x)d\xi = -\frac{1}{2}\ln(x^2+y^2) - \gamma \ , \quad I_{10} \equiv \int_0^\infty \frac{1}{\xi} e^{-\frac{y(1+\ell^2\xi^2)^{1/2}}{\ell}}\sin(\xi x)d\xi \ ,$$

$$I_{11} \equiv \int_0^\infty \frac{(1+\ell^2\xi^2)^{1/2}}{\xi} e^{-\frac{y(1+\ell^2\xi^2)^{1/2}}{\ell}}\sin(\xi x)d\xi = -\ell\frac{\partial}{\partial y}(I_{10}) \ ,$$

where $\gamma$ is Euler's constant, $r = (x^2 + y^2)^{1/2}$, and $K_i(x/\ell)$ is the i$^{th}$ order modified Bessel function of the second kind. Integrals $I_m$ ($m = 1,2,...,9$) were obtained in closed form but integrals $I_{10}$ and $I_{11}$ have to be evaluated numerically. In view of the above, the displacement field reads finally

$$u_x(x,y) = \frac{b(1-2v)}{4\pi(1-v)}\ln(r) + \frac{b(y^2-x^2)}{8\pi(1-v)r^2} + \frac{b(y^2-x^2)}{2\pi r^2}\left[K_2(r/\ell) - \frac{2\ell^2}{r^2}\right] + \frac{b}{2\pi}K_0(r/\ell)$$

$$-\frac{\Omega \ell^2 x}{\pi r^2} + \frac{\Omega \ell}{\pi}I_{11} + \frac{\Omega y}{4}, \tag{A3}$$

$$u_y(x,y) = \frac{b}{2\pi}\theta - \frac{bxy}{4\pi(1-v)r^2} - \frac{bxy}{\pi r^2}\left[K_2(r/\ell) - \frac{2\ell^2}{r^2}\right] + \frac{\Omega y}{2\pi}\left[K_2(r/\ell) - \frac{2\ell^2}{r^2}\right]$$

$$-\frac{\Omega y}{2\pi}K_0(r/\ell) - \frac{\Omega x}{4}. \tag{A4}$$

Further, the rotation is given as

$$\omega(x,y) = \frac{by}{4\pi\ell^2}\left[K_2(r/\ell) - \frac{2\ell^2}{r^2}\right] - \frac{by}{4\pi\ell^2}K_0(r/\ell) + \frac{\Omega}{2\pi}I_{10} - \frac{\Omega}{4}. \tag{A5}$$

It is noted that the rotation in (A5) is discontinuous at $y = 0$ due to the integral $I_{10}$. To show this, we expand the integrand of $I_{10}$ in series as $\xi \to \infty$, i.e.

$$(1/\xi)e^{-\frac{y(1+\ell^2\xi^2)^{1/2}}{\ell}} = e^{-y\xi}\left[(1/\xi) - y/(2\xi^2\ell^2) + y^2/(8\ell^4\xi^3) + ...\right]. \tag{A6}$$

Then, the inverse Fourier sine transform of the first term in the above series is given by $I_1$ and it is clearly discontinuous at $y = 0$ while all the other terms do not contribute at $y = 0$. On the other hand, it can readily be seen that the part of the normal displacement in (A4) due to the constrained wedge disclination is everywhere continuous. Thus, it is apparent that the discontinuity in the rotation *does not affect* the normal displacement in the case of a constrained wedge disclination. Finally, we note that a rigid body translation $b/4$ and a rigid body rotation $-\Omega/4$ have been added in (A4) and (A5), respectively, in order to have zero normal displacement and rotation at $y = 0^+$, $x > 0$.

The stress and couple-stress field can now be obtained using (A3)-(A5). In particular, we have[#]

$$m_{yz}(x,y) = -\frac{\mu b}{\pi}\frac{(x^2-y^2)}{r^2}\left[\frac{2\ell^2}{r^2}-K_2\left(\frac{r}{\ell}\right)\right] - \frac{\mu b}{\pi}K_0\left(\frac{r}{\ell}\right), \tag{A7}$$

$$m_{xz}(x,y) = \frac{2\mu b}{\pi}\frac{xy}{r^2}\left[\frac{2\ell^2}{r^2}-K_2\left(\frac{r}{\ell}\right)\right], \tag{A8}$$

$$\sigma_{yy}(x,y) = \frac{\mu b\, x}{2\pi(1-\nu)}\frac{(3y^2+x^2)}{r^4} - \frac{2\mu b\, x}{\pi}\frac{(3y^2-x^2)}{r^4}\left[\frac{2\ell^2}{r^2}-K_2\left(\frac{r}{\ell}\right)\right] \\ + \frac{\mu b}{\pi\ell^2}\frac{xy^2}{r^2}\left[K_2\left(\frac{r}{\ell}\right)-K_0\left(\frac{r}{\ell}\right)\right], \tag{A9}$$

$$\sigma_{xx}(x,y) = \frac{\mu b\, x}{2\pi(1-\nu)}\frac{(x^2-y^2)}{r^4} + \frac{2\mu b\, x}{\pi}\frac{(3y^2-x^2)}{r^4}\left[\frac{2\ell^2}{r^2}-K_2\left(\frac{r}{\ell}\right)\right] \\ - \frac{\mu b}{\pi\ell^2}\frac{xy^2}{r^2}\left[K_2\left(\frac{r}{\ell}\right)-K_0\left(\frac{r}{\ell}\right)\right], \tag{A10}$$

$$\sigma_{yx}(x,y) = \frac{\mu b}{2\pi(1-\nu)}\frac{y(x^2-y^2)}{r^4} - \frac{2\mu b\, y}{\pi}\frac{(3x^2-y^2)}{r^4}\left[\frac{2\ell^2}{r^2}-K_2\left(\frac{r}{\ell}\right)\right] \\ + \frac{\mu b}{\pi\ell^2}\frac{yx^2}{r^2}\left[K_2\left(\frac{r}{\ell}\right)-K_0\left(\frac{r}{\ell}\right)\right], \tag{A11}$$

$$\sigma_{xy}(x,y) = \sigma_{yx}(x,y) - 4\mu\ell^2\nabla^2\omega. \tag{A12}$$

The above expressions are the *full-field* Green's functions for the mode I problem. Further, it is worth noting that when $y=0$ (imagined crack-line), the integral $I_{11}$ can be evaluated analytically in the *finite-part* sense (see e.g. Zemanian, 1965; Roos, 1969) as

$$\int_0^\infty \frac{(1+\ell^2\xi^2)^{1/2}}{\xi}\sin(\xi x)d\xi = -\frac{1}{4}\mathrm{sgn}(x)\cdot G^{2,1}_{1,3}\left(\frac{x^2}{4\ell^2}\left|\begin{array}{c}1\\-1/2,\,1/2,\,0\end{array}\right.\right), \tag{A13}$$

where $G^{a,b}_{c,d}(\ )$ is the MeijerG function. Thus, the Green's functions for the mode I crack problem can be obtained in closed form (Eqs. (60) and (61) of the main body of the paper).

---

[#] Note that in [Gourgiotis, P.A., Georgiadis, H.G., 2008. An approach based on distributed dislocations and disclinations for crack problems in couple-stress elasticity. *Int. J. Solids Struct.* 45, 5521-5539] the expressions for ($\sigma_{xx}$, $\sigma_{yy}$, $\sigma_{yx}$) (see Eqs. (A9)-(A11) in that paper) contained few misprints that do not, however, affect their final results in any way.

**Appendix B: Derivation of the quadrature for the integral with the logarithmic kernel**

Consider the *weakly singular* integral

$$I(t) = \int_{-1}^{1} f(s)\cdot(1-s^2)^{-1/2}\ln(p|t-s|)ds, \tag{B1}$$

where $f(s)$ is a continuous bounded function in $|s|\leq 1$ and $p$ is a positive constant. Now, (B1) can be written as follows (Theocaris et al., 1980)

$$I(t) = \int_{-1}^{1} R(s,t)\cdot(1-s^2)^{-1/2}ds + f(t)\int_{-1}^{1}(1-s^2)^{-1/2}\ln(p|t-s|)ds, \tag{B2}$$

where $R(s,t) = [f(s)-f(t)]\cdot\ln(p|t-s|)$ is a bounded function in the interval $-1\leq (s,t)\leq 1$. The Gauss-Chebyshev quadrature rule is employed for the evaluation of the first integral in (B2), whereas the second integral can be evaluated in closed form. In light of the above, we obtain

$$I(t) \cong \frac{\pi}{n}\sum_{i=1}^{n} R(s_i,t) + \pi\ln(p/2)f(t) = \frac{\pi}{n}\sum_{i=1}^{n} f(s_i)\cdot\ln(p|t-s_i|) + f(t)\cdot G_n(t), \tag{B3}$$

where $G_n(t) = -\frac{\pi}{n}\sum_{i=1}^{n}\ln(p|t-s_i|) + \pi\ln(p/2)$ and the integration points are given as the zeros of the Chebyshev polynomial $T_n(s)$, i.e. $s_i = \cos[(2i-1)\pi/2n]$, $i=1,2,\ldots,n$.

One further step is needed which would lead to the evaluation of the RHS of (B3) only at $n$ points $s_i : T_n(s_i) = 0$. This can be done with the aid of the Lagrange interpolation formula, which is exact within the class of polynomials chosen to represent $f(t)$, i.e.

$$f(t) = \sum_{i=1}^{n} \frac{T_n(t)}{T_n'(s_i) \cdot (t - s_i)} f(s_i) . \tag{B4}$$

Integral $I(t)$ takes now the discretized form

$$I(t) \cong \frac{\pi}{n} \sum_{i=1}^{n} f(s_i) \cdot \ln(p|t - s_i|) + G_n(t) \cdot T_n(t) \sum_{i=1}^{n} \frac{f(s_i)}{T_n'(s_i) \cdot (t - s_i)} , \quad s_i = \cos[(2i - 1)\pi/2n] . \tag{B5}$$

Finally, we note that, in the system of singular integral equations (73), *prescribed* collocation points in (B5) have been chosen, i.e. the zeros of the Chebyshev polynomial $U_{n-1}(t)$: $t_k = \cos(k\pi/n)$, $k = 1, 2, ..., n - 1$, in order for us to be consistent with the numerical quadrature that was employed for the Cauchy-type singular integrals. We note that another type of quadrature using *arbitrary* collocation points for the solution of integral equations with logarithmic singularities was proposed by Chrysakis and Tsamasphyros (1992).